A predictive bond order model connects covalent superconductivity, including cuprates, oxypnictides or HfN based materials.

H. Oesterreicher, Department of Chemistry, UCSD, La Jolla, CA 92093-0506

**Abstract**

A phenomenological bond order model predicting doping curves [Tc vs doped charge, c] of cuprates is found to have wider applicability. In this model Tc is dependent on the density of electronic pair crystals [EC] in a covalently bonding layer structure and on a layer Isolation factor, f [ECI model]. Characteristic doping curve events such as optima are correlated with a select number of EC with pair repeats corresponding to multiples of lattice parameters such as $c=2/3x4=0.167$, where 3x4 represent pair periodicity of $3a_0$ and $4b_0$. At these EC all doped charge is converted into pairs according to $c=2n_p$. For Tc prediction one writes $Tc=2n_p f T^e$, where $T^e$=600K and 300K as empirical constants for hole and electron doping. Doping curves for $YBa_2Cu_3O_y$ with their sharp optima or kinks, separated by near linear ranges, or Tc plateaus on different preparations, can express the stability of special EC. Examples are $c=0.22=2/3^2$ for the sharp optimum or the Tc=90K plateau, and $c=0.17$ for the 60K plateau, the latter depicting also the optimum for other systems such as $La_{2-w}Sr_wCuO_4$. For oxypnictides $R[O_{1-x}F_x]FeAs$ one writes $Tc=300x$ and expects EC with optimal dopings at $x=0.11$ and $0.17$, as corroborated experimentally. Other examples include HfN derivatives such as $Li_{0.17}HfNCl$.

**Introduction**

New materials and general considerations:

Iron-oxypnictide superconductors were discovered on LaOFeP (Tc = 5K) [1] and LaOFeAs (Tc = 26K at x=0.11) [2], followed by subsequent development of materials with higher Tc's containing rare earth (R) elements, such as Ce, Nd, Sm [3-6] instead of La. For Sm at optimal x=0.17, Tc=55K. Carriers are doped by (O, F) or (Gd,Th) substitutions [[7] Tc=56.5 K] as well as high-pressure oxygen synthesis [8]. Generally one notices that the Tc value increase with decreasing lattice constants in $RO_{1-x}F_xFeAs$ at 'optimal' doping level. Unfortunately, the parent compounds ROFeAs with the ZrCuSiAs-type structure are limited to R=La, Ce, Pr, Nd, Sm and Gd [9]. Muon spin relaxation (μSR) studies [10] demonstrate several generic features common to the iron-oxypnictides and cuprate systems. They include long-range commensurate antiferromagnetism of the undoped parent compounds, incommensurate or stripe magnetism of lightly doped systems near the border of magnetic and superconducting phases, and scaling of the superfluid density and Tc.

Charge ordering effects, as a common basis for these superconductors, are ubiquitous in transition metal oxides [11-13]. Not surprisingly, mounting experimental [14-17] evidence exists for the operation of such effects also in high $T_c$ superconductivity. Microscopic experiment [STM, ARPES] [14-17] further emphasizes charge ordering. Such a view was previously explored in terms of the t-j [band-exchange] model or the direct involvement of doped charge in stripes [18-20] of single holes. It was also anticipated within an alternative phenomenological model [21-28] where superconducting electronic 'plaid' crystals [EC] grow out of stripe phases [29].

The experimental indications for the involvement of EC in high $T_c$ superconductivity in cuprates renewed the question for predictive formalisms concerning Tc. In fact, the tendencies for non stoichiometric bonding to create special ordered structures were considered early on [21-28] and patterns of superexchange pairs were proposed, with their density directly proportional to Tc. Ranges in the doping curves [Tc versus doped charge], leading up to the optimum correspond to conditions where are all holes are transformed into pairs, denoted by $h_p$, or in simple context by h. When not distinguishing between hole and electron doping, doped charge concentration is denoted by c and the region where all doped charge is converted into pairs is $c=2n_p$. Charge-lattice commensurabilities explained the trends to a select number of optimal doping concentrations. These EC, also refered to as bond orders [BO], in the cuprates include $h_p$=0.167, 0.22 or 0.25, as later verified by Knight shift [30]. These "magic" numbers have been shown to correspond to the charge-lattice commensurability requirements of EC and are demonstrated in ARPES. When combined with formalisms to account for the advantageous effects of layer isolation, they allow for a predictive formalism in the electronic crystal isolation, or ECI model. With the advent of new classes of high Tc materials, such as the oxypnictides, the usefulness of the ECI model should be further testable.

Introduction to the ECI model:

The history of phenomenological laws for the cuprates suffered from a flaw that prohibited earlier appreciation for the involvement of charge ordering. This flaw concerns the lack of recognition of the true nature of doping curves as containing, in their uncompromised examples, 'sharp' features such as optima at select charge ordering related hole concentrations. Historically, the first doping curve were obtained on $La_{2-w}Sr_wCuO_4$ and showed a broad parabolic appearance with an optimum near $h_p$=0.16 [$La_{1.84}Sr_{0.16}CuO_4$]. This was taken to represent the prototype behavior, reminding of an involvement of bands, even though one of the materials studied next, $YBa_2Cu_3O_y$, did not depict itself in anything like it [refs in 28]. Rather it can show marked transitions between more or less linear regions, including a sharp optimum corresponding to $h_p$=0.22 determined by Knight shift [30]. On special preparations, Tc-plateaus were observed. The difference in doping curve shape between $La_{2-w}Sr_wCuO_4$ and $YBa_2Cu_3O_y$ has to do with the proximity of the doping to the "plane" in the first system, corrupting the inherent charge ordering signatures that can be so clearly visible in the latter.

In the ECI model, local super-exchange pairs form electronic crystal arrangements in superconducting pair plaids [21, 24-28], that can be created out of non-superconducting stripes [29] by rearrangement [28]. They represent unique selected charge–lattice lock-in patterns, explaining trends to a select number of characteristic optimal hole numbers. Accordingly, the period of the electronic crystal is related to hole-concentration, which corresponds to the 2 charges of the pair, taken per number of atoms in a tile as defined by its period. This period Pab, where 'a' and 'b' are multiples of lattice parameter [e.g.P3x3 for $3a_0x3b_0$ also depicted as h=0.222=$2/3^2$, where h is now understood to stand for $h_p$. Other example are h=2/3x4=0.167 or h=0.125=2/4x4]. These 'magic' fractions, deduced from fundamental considerations [21, 23], produce selected charge-lattice lock-in patterns at h= 0.125, 0.167, 0.20, 0.22 or 0.25, which are the basis of the characteristic dopant concentrations ubiquitously encountered in experiment. In fact several of the predicted EC [such as 2/4x4, 2/3x4 or 2/3x3] can be clearly seen within local regions with ARPES corroborating their earlier prediction on fundamental considerations. In specially prepared $YBa_2Cu_3O_y$, h=0.222 and h=0.167 are considered to correspond to the Tc~90 and 60K plateau respectively. These optimal hole concentrations give the number of pairs $n_p$=1/Pab. The latter also represents the elastic correlation energy [for the linear 'source' region, where all holes are transformed into pairs $2n_p=h_p$]. This series of optimal charge lattice lock-in patterns and the resulting Tc calculations are successfully born out by Knight shift [30] and fit into the predictions of ECI.

A cornerstone of cuprate phenomenology is the sensitivity of Tc to the density of this layered electronic crystal and to proximity and nature of its c-axis coordination as reflected in quantitative algorithms involving bond valence in the c-direction. In the Electronic Crystal Isolation model then

$$T_c=2f600n_p=2f600/Pab= fh_p600$$

Here, $h_p$ represents the holes that occur as pairs, the apical factor f stands for the diminished correlation of the plane configurations, when the apical system interacts with the plane and 600 an empirical factor $T^e$. f depends on the bond valence [28] in the c-direction or, more specifically, the difference in bond valence between superconductor and undoped parent compound. This parameter can be seen as related to the hybridization out of the plane [31]. Where relevant data are unavailable, f can be approximated by $f_a$ =1, 2/3, ½ for apical layer coordination of 0,1 and 2 [$CuO_2$, $CuO_3$, $CuO_4$] respectively. The trends for $f_a$ and Pab to characteristic values organize cuprates into a level structure corresponding to 'musical' $T_c$ families. Empirical rules allow for compound specific theoretical $T_c$ and doping curve [Tc versus h] predictions purely on structural parameters.

**Results**

The ECI procedure outlined above holds for representative cuprate high $T_c$ compound families apparently so far without exceptions. The ECI concepts are used in the following to explain selected aspects of high $T_c$ superconductivity in oxypnictides and relate them to the cuprates. A sampling of $T_c$ mainly on the $f_a$ approximation and in one case demonstrating the f algorithm for cuprates is given in TABLE 1 in order to compare it with calculations of oxypnictides in this study. A first part of this study identifies the Tc- rules for oxypnictides as empirical Tc=300x and compares it to the cuprates with Tcmax= 600h, the maximum attainable with f=1. A second part interprets this finding in a more general sense, suggesting a direct equivalence of 2x and h. Accordingly for electron doped oxypnictides Tc=x2$f_a$300. This indicates a fundamental equivalence of phenomenology in both cases. In a second part this idea is extended to other layer compounds based on electronegative elements.

Tc- rules and predicted charge ordering for oxypnictides:

Structural analysis shows that the FeAs layer has a buckled appearance. When presented on the plane, the slightly more electronegative component As displays a centered lattice, as does Fe. The centers of the doped negative charge are residing in FeAs bonds. We start with an EC corresponding to $3a_0 \times 3b_0$ which corresponds with the optimum in $YBa_2Cu_3O_{y\sim6.9}$. Due to the lattice centering by As, we find that 2x=0.222 corresponds to h=0.222 for an EC with similar lattice periodicity. For oxypnictides this corresponds to 2/18As=0.11/As=x, while for cuprates it is 2/9Cu= 0.22/Cu=h. Structural analysis therefore shows that for comparable EC conditions based on unit cell, 2x corresponds to h. For direct comparison one can represent this by a formulation $La_2[O_{2-2x}F_{2x}]Fe_2As_2$.

Empirically we find for oxypnictides Tc=300x [TABLE 1]. This, for $La_2[O_{2-2x}F_{2x}]Fe_2As_2$, with 2x=0.22, gives calculated Tc=33$^\circ$K, compared to observed 26$^\circ$K which shows optimum indeed at x=0.11. The observed Tc plateau for lower x reminds of the one for $YBa_2Cu_3O_{6.9}$ with Tc~90K. In analogy to $YBa_2Cu_3O_y$ one could expect for the oxypnictides another characteristic doping concentration with $3a_0 \times 4b_0$ periodicity or h=2x=0.167=2/3x4 as in $YBa_2Cu_3O_{6.6}$. For the oxypnictides this apparently lies hidden in the Tc plateau that comes to an end for 2x<0.083 similar to the end of the linear region in Tc vs. h in $YBa_2Cu_3O_{6.3}$. It does not create its own plateau as with the Tc~60K one of $YBa_2Cu_3O_{6.6}$. Tc-onset is expected and found near 2x=h~0.05 in both cases, along arguments of competition with stripe structures.

However, a further densification of EC with $2/3^2$ is in principle possible, yielding h=e=0.333=2/3x2. For the oxypnictides this gives x=0.167 and we suggest here that this EC is the basis of the higher T~55$^\circ$K obtained for x=0.17 found in experiment with the Sm-analog. For the expected EC with x=0.167 one calculates Tc=50$^\circ$K. It should be noted that an EC with corresponding optimal hole concentration of h=0.33 has so far not been found for the cuprates, indicating an expanded phenomenology with the oxypnictides. An even denser EC can be expected on centering of EC according to $4/3^2$, yielding h=0.444, or with arrangements such as h=2x=$2/2^2$=0.5. As with the cuprates, this charge densification can be seen as a relative increase in the population of secondary charge channels [28].

Fundamental equivalence of phenomenology in 'covalent' superconductivity:

We search now for the deeper relationships between the empirical formulas. The simple empirical ECI relation for cuprates is Tc=h$f_a$600, while we observe Tc=300x for oxypnictides in Table 1. At face value we can take $f_a$=1/2 for the oxypnictides, due to the twofold coordination of the AsFe layer. In addition we compare here electron and hole doped materials. In the available small number of cuprate examples with reliable dopant counts, indications are that electron doped materials have Tc considerably reduced [by about 1/2] compared to values of the hole doped case [25 for refs]. In order to erect an orienting frame, we assume that the relevant formulation can be approximated by Tc=$e_p$f300 corresponding to Tc=x2$f_a$300.

We compare cuprates and oxypnictides. Accordingly $HgBa_2Ca_2Cu_3O_{8+x}$ [OP= outer planes] can serve as an example for approaching Tcmax=150K at h~0.25. With f=0.88, the calculated Tc=132$^\circ$K and the observed value is Tc=133$^\circ$K. Similar values are obtained for the inner plane at h~0.22 and $f_a$=1. h~0.22 is also characteristic for optimum in $YBa_2Cu_3O_y$ but here $f_a$=2/3. Full doping curves are similar for this system and the exemplary $La[O_{1-x}F_x]FeAs$. Both show tendencies to Tc plateaus in experiment although the former can also obtain with an extended near linear region up to the optimum. At fictitious h=0.5 and f=1 one calculates Tc=300K. It is possible however, that the oxypnictides will not accept hole doping.

Suitable compounds for direct comparison are $Sm[O_{1x}F_x]FeAs$ and $La_{1.84}Sr_{0.16}CuO_4$. Both have a doubly coordinated superconducting layer and therefore, in a simple definition, identical $f_a=1/2$. However, as an electron doped material $Sm[O_{1x}F_x]FeAs$ should have half the Tc compared the hole doped cuprate and would display comparable Tc only if its doped charge were doubled, that is at 2x=0.33. This is indeed observed. $La_{1.84}Sr_{0.16}CuO_4$ has observed Tc=38°K around h=0.16, and a calculated Tc=48$^c$K for x=0.16 [$f_a$ is actually <1/2 due to particularly close apical approach, further reducing Tc]. This is compared to the observed Tc=55°K for Sm analog oxypnictides at 2x=0.33, for which we calculate Tc=50$^c$K. There exists therefore a direct relationship between both families.

Another pair for comparison is $La[O_{1-x}F_x]FeAs$ and $YBa_2Cu_3O_{6.9}$. Both have a similar EC with $h=e=2/3^2=0.22$. To first approximation $YBa_2Cu_3O_{6.9}$ should have double the Tc as a hole doped compound. The yet higher Tc is related to the advantageous $f_a=2/3$ of only single layer coordination with apical O.

Similar comparisons can be made for other compounds in Table 1. The examples of higher Tc in the cuprates originate from the further reduction in apical interaction such as in materials based on Hg, or the loss of significant apical interaction altogether in multilayer compounds. This can lead to yet higher optimal h, for which no counterparts [e.g. x=0.25] have yet been found in the oxypnictides. On this similarity alone one can expect yet higher Tc values in the oxypnictides.

It is interesting that an infinite layer compound $BaFe_{2-2x}Co_{2x}As_2$ with 2x=0.2 (Tc = 22K, ref 32) can also roughly be accounted on in ICE. Assuming Co to donate one electron to the $Fe_{2-2x}Co_{2x}As_2$ system one calculates Tc=300x= 30K.

The experience outlined above is condensed into the perhaps most general relationship between cuprates and oxypnictides by involving the definition of charge period P. Considering the difference in basic counting of charge we write $T_c=hf_a600 =2f_a600/Pab$ and $T_c=xf_a300 =f_a600/Pab$, for hole doped cuprates and electron doped oxypnictides respectively. This holds for the region where all doped charge is converted into pairs.

The concepts of characteristic EC on covalent layer structures and the proportionality of Tc to their pair density, modified by an isolation factor, can be expanded to other materials. We proceed here under the simplifying assumption of $f_a=1/2$. For reference below we calculate Tc=25$^c$K at $e_p=0.16=4/5^2$ and Tc=38$^c$K at $e_p=0.250=4/4x4$, with double these Tc for hole doping. Compounds with HfN layers like $Li_{0.17}HfNCl$ with observed Tc=25°K can accordingly be derived from $e_p=0.16=4/5^2$ [or $e_p=0.167=2/3x4$]. This probably also holds for doped HfN [Tc=18°K], even though the doping is here more directly corrupting the layer. It is also suggested for $PbMo_6S_8$ [Tc=18°K] [for refs see 25]. One can also attempt to make assignments for the higher EC and Tc family with $e_p=0.25$ and Tc=38$^c$K. Examples are $MgB_2$ [Tc=38°K], where EC formation would involve the B layer through selfdoping from Mg, or C-S composites [Tc=35°K].

**Discussion**

While the Tc data on oxypnictides are presently limited, it is encouraging that a basic application of the concepts of real space pair ordering in covalent layers and the importance of layer isolation in ECI holds. Again one finds indications for the canonical EC or derivatives thereof in the optimal dopings such as x=0.17 for which a corresponding value is found repeatedly in the cuprates. It would appear that characteristic EC formation and with it superconductivity is a natural consequence of doping of any relatively isolated layered covalent bond system. An example for the adherence to one of the more prevalent characteristic dopant charge numbers is found in $Li_{0.17}HfNCl$ where one encounters similar covalent layers [HfN]. Accordingly, signs are that the concept of charge patterning on covalent substructures is further extendible to a wider range of systems. The indication for success of a generalized phenomenological predictive scheme, one that roughly orders the various compound classes, forms presently a challenge to theory.

The importance of layer isolation comes into play in a variety of ways. It fine-tunes Tc at a specific doping. It also determines the magnitude of the optimal doping. In fact, strong bonding out of the covalent layer may prove to represent the primary reason why many, at first sight suitable systems such as oxypnictides based on NiAs, either fail to superconduct or show low Tc. $PrBa_2Cu_3O_{6.9}$ can serve as an example of lack of superconductivity due to anomalies in axial ratios on special preparations [28]. This prediction can be considered a particular success of ECI. In fact, the intent of ECI is not so much directed at high accuracy in Tc prediction but in showing the overarching Tc rules and the expectation for new materials.

One point of ECI therefore is to indicate that the competition of pairs with other charge arrangements such as stripes of singles is shifted in favor of the former in cases of high isolation factor f. This EC formation of paired resonant 5 center bonds appears to be a fundamental response of an isolated doped covalent layer in a special range of dopant concentration. This fundamentally new resonant bonding arrangement along specified charge channels appears to be the primary condition for the existence of high Tc-superconductivity. It forms perhaps the best operational distinction to "non-isolated layer" metallic superconducting systems such as Pb.

It is also gratifying that the general energy scales in the prediction of Tc appear related. However, while this indicates that the concepts of geometrically dictated characteristic EC at specified charge concentrations are general to all high Tc materials and express themselves in a quantitatively comparable manner, these general energy scales open many fundamental questions. Do they indicate that parameters such as the electrostatic repulsion amongst pairs are of fundamental importance.

It is satisfying that the two prominent optimal dopant concentrations found so far for oxypnictides, namely x=0.11 and 0.17, bear out the concept of a succession of favorable EC in various pair plaid patterns. Indications are that these patterns are related in the cuprates and oxypnictides. The optimal EC, corresponding to 3ax3b lattice periodicity, with 2x=0.22/2As and h=0.22/Cu in two selected cases present convincing examples of this concept. Also Tc onsets at 2x=h~0.05 testify to similar origins. It should be noted however that the corresponding optimal hole concentration of 2x=0.33, indicated for the oxypnictides, has so far not been reached for the cuprates. This suggests an expanded phenomenology for the oxypnictides. It shows promise for further increases of Tc involving more concentrated EC corresponding to 2x= 0.444 and 0.5 for both classes. It also should allow for an understanding of the spatial requirements for dense EC. Opportunities for further Tc tailoring of the oxypnictides lie in a general chemical engineering of increased relative plane isolation. Ways to increase optimal doping through increasing of f have been reviewed for cuprates [28] and should be applicable to the oxypnictides as well.

Local models such as ECI answer to an impressive array on the wish list for high $T_c$ superconductivity. This array ranges from quantitative doping curve predictions to pictorial representation in pair crystals. It includes requisites for its generation, such as layers of covalent bonds with high exchange fostered by high isolation from adjacent layers. ECI as based on elasticity and exchange is qualitatively distinct in assumptions and conclusions, e.g. from the somewhat related t-j model. As an example, a version of the latter considers the initial $T_c$ Rise region to contain superconducting stripes that become 2-dimensional at higher $T_c$. By comparison ECI considers only pair strands to superconduct. It contains direct proportionality of $n_p$ and $T_c$ and straightforward mechanisms for coexistence of single holes and pairs and their mutual transformations. The initial strong rise of Tc marks the region of transformation of stripes of singles into plaids of pairs. However, a convergence and synergy amongst these approaches appears possible.

Table 1 Charge order periodicity and resulting characteristic hole numbers and Tc=600$f_a$h or Tc=300x for the main types of bond patterns. They are the basis of a 'musical' $T_c$ level scheme. $T_{cmax}$ would be obtainable if $f_a$=1. OP and IP stand for outer and inner plane respectively. $^o$ stands for observed [3], $^c$ for calculated on $T_c$= $f_a$ $T_{cmax}$. h, e stand for experimental values. For HgBa$_2$CaCu$_3$O$_{8+x}$ OP, $f_a$ =0.88>2/3 with $T_c$=132$^c$K is due to its unusually large distance to the apical O. Materials with anomalous low c/a can be nonsuperconducting such as PrBa$_2$Cu$_3$O$_7$. Reliable hole concentration in the cuprates have been determined by Knight shift [30].

| $2x=n/Pab$ | $T_{cmax}[K]=600h$ | Examples | h, e | $f_a$ | $T_c[K]$ |
|---|---|---|---|---|---|
| Oxypnictides | | | | | |
| 0.500=2/2x2 | 300 | x=0.25 | | ½ | $75^c$ |
| 0.444=4/3x3 | 267 | x=0.22 | | ½ | $67^c$ |
| 0.333=2/2x3 | 200 | Sm[$O_{1-x}F_x$]FeAs, x=0.17 | 0.33 | ½ | $50^c$ [x=0.167], $55^o$ |
| 0.222=2/3x3 | 133 | La[$O_{1-x}F_x$]FeAs, x=0.11 | 0.22 | ½ | $33^c$, $26^o$ |
| h=n/Pab | | | | | |
| Cuprates | | | | | |
| 0.444=4/3x3 | 267 | | | | |
| 0.333=4/3x4 | 200 | | | | |
| 0.250=4/4x4 | 150 | $Bi_2Sr_2CaCu_2O_{8.25}$ 'Tet' | 0.25 | 2/3 | $100^c$, $92^o$ |
| 0.250=4/4x4 | 150 | $HgBa_2Ca_2Cu_3O_{8+x}$ [OP] Tet | 0.25 | 2/3<f=0.88 | $132^c$, $133^o$ |
| 0.222=4/3x6 | 133 | $YBa_2Cu_3O_{6.95}$ Ort | 0.22 | 2/3<f | $89^c$, $95^o$ |
| 0.222=2/3x3 | 133 | $HgBa_2Ca_2Cu_3O_{8+x}$ [IP] Tet | 0.22 | 1 | $133^c$, $133^o$ |
| 0.200=4/4x5 | 120 | $YBa_2Cu_4O_8$ Ort | 0.19 | 2/3 | $80^c$, $80^o$ |
| 0.167=2/3x4 | 100 | $YBa_2Cu_3O_{6.7}$ Ort | 0.16 | 2/3 | $66^c$, $64^o$ |
| 0.160=4/5x5 | 96 | $La_{1.84}Sr_{0.16}CuO_4$ | [0.16] | ½ >f | $48^c$, $38^o$ |
| 0.125=2/4x4 | 75 | $GdBa_2RuCu_2O_8$ Tet | [0.125] | 2/3 | $50^c$, $48^o$ |
| 0.125=1/2x4 | - | 'Tranquada stripes' | | | |
| 0.0833=2/4x6 | 50 | kink $YBa_2Cu_3O_{6.3}$ | | 2/3 | $33^c$, $30^o$ |

REFERENCES

1. Y. Kamihara Y. Kamihara, T. Watanabe, M. Hirano, H. Hosono J. Am. Chem. Soc. 128 (2006) 10012.
2. Y. Kamihara, T. Watanabe, M. Hirano, H. Hosono, J. Am. Chem. Soc. 130 (2008) 3296.
3. X.H. Chen T. Wu, G. Wu, R. H. Liu, H. Chen and D. F. Fang, Cond-mat: arXiv:0803.3603 (2008)
4. G.F. Chen, Z. Li, D. Wu, G. Li, W. Z. Hu, J. Dong, P. Zheng, J. L. Luo, N. L. Wang,., Cond-mat arXiv:0803.3790 (2008).
5. Zhi-An Ren, Jie Yang, Wei Lu, Wei Yi, Guang-Can Che, Xiao-Li Dong, Li-Ling Sun, Zhong-Xian Zhao, Cond-mat arXiv:0803.4283 (2008).
6. H.H. Wen, G. Mu, L. Fang, H. Yang, X. Zhu, Europhys.Lett. 82 (2008) 17009.
7. Cao Wang, Linjun Li, Shun Chi, Zengwei Zhu, Zhi Ren, Yuke Li, Yuetao Wang, Xiao Lin, Yongkang Luo, Xiangfan Xu, Guanghan Cao_ and Zhu'an Xu, Cond-mat arXiv:0804.4290,doi:10.1038/nature06972.
8. H. Takahashi, K. Igawa, K. Arii, Y. Kamihara, M. Hirano, and H. Hosono, Nature, doi:10.1038/nature06972.
9. P. Quebe, L. J. Terbuchte, and W. Jeitschko, Journal of Alloys and Compounds 302, 70 (2000).
10. J. P. Carlo, Y. J. Uemura,T. Goko, G. J. MacDougall, J. A. Rodriguez, W. Yu, G. M. Luke, Pengcheng Dai, N. Shannon, S. Miyasaka, S. Suzuki, S. Tajima, G. F. Chen, W. Z. Hu, J. L. Luo, and N. L. Wang, Cond-mat arXiv:0805.2186v1
11. A.R. Bishop. Current Opinion in Solid State and Material Science 2, 244 (1997.
12. J.B. Goodenough and J.S. Zhou. In: Structure and Bonding 98, 17 (2001).
13. C.Park, R.L.Snyder J. Am. Cer. Soc.78,3171 (1995)
14. J.E. Hoffman, E.W. Hudson, K.M. Lang, VV. Madharan, H. Eisaki, S. Uchida, J.C. Davis. Science 295, 466 (2002).
15. H.A. Mook, P. Dai, F. Dogan. Phys. Rev. Letters 88, 97004 (2002).
16. T. Hanagurii, Nature, 430, 1001 (2004)
17. K. M. Shen, F. Ronning,D. H. Lu, F. Baumberger, N. J. C. Ingle, W. S. Lee, W. Meevasana, Y. Kohsaka, M. Azuma, M. Takano,H. Takagi, Z.-X. Shen, Science, 307, 901-904 [2005]
18. Y. Zhang, E. Demler, S. Sachdov. Phys Rev. B 66, 94501 (2002).

19. J. Zhu, I. Martin, R.A. Bishop Phys Rev. Letters 89, 67003 (2002).
20. M. Voita Phys Rev. B, 66, 104505 (2002).
21. H. Oesterreicher, J. Solid State Chemistry, 158,139 (2001)
22. J.B. Goodenough.  Europhys Letters 57, 550 (2002).
23. H. Oesterreicher, J. Superconductivity 16, 507, (2003)
24. H. Oesterreicher, J. of Alloys and Compounds, 366, 1, (2003)
25. H. Oesterreicher, J. Superconductivity 17, 439, (2004) and 18, 509 [2005]
26. H. Oesterreicher Solid State Communications, 137, 235-240 [2006]
27. H. Oesterreicher, AIP Proc. 24$^{th}$ international conference on low temperature physics, p573 [2006]
28. H. Oesterreicher Solid State Communications, 142, 583-586 [2007]
29. J.M. Tranquada, B.J. Sternlieb, J.D. Axe, Y. Nakamura and S. Uchida.  Nature 375, 561-563 (1995).
30. H. Kotegawa, Y. Tokunaga, K. Ishida, G.-Q. Zhang, Y. Kitaoka, H. Kito, A. Iyo, K. Tokiwa, T. Watanabe, and H. Ihara.  Phys. Review B, 64, 064515 [2001]
31. L.F. Feiner, M. Grilli, C. DiCastro, Phys Rev. B 45, 10647 (1992).
32. F. L. Ning, K. Ahilan, T. Imai, A. S. Sefat, R. Jin, M. A. McGuire, B. C. Sales and D. Mandrus arXiv:0808.1420 (August 2008)